\documentclass[12pt]{article}
\begin{document}

\title{\textbf{VECTOR CONSTANTS OF MOTION FOR TIME-DEPENDENT KEPLER AND ISOTROPIC
HARMONIC OSCILLATOR POTENTIALS}}
\author{O.M. Ritter$^{\star }$ \\
Departamento de F\'{\i}sica\break
\\
Universidade Federal de Santa Catarina\break 
\\
Trindade \break 
\\
88040-900 Florian\'{o}polis SC, Brasil \and F.C. Santos$^{\dagger}$ and A.C.
Tort$^{\ddagger}$ \\
Instituto de F\'{\i}sica\break 
\\
Universidade Federal do Rio de Janeiro\break 
\\
Cidade Universit\'aria - Ilha do Fund\~ao - Caixa Postal 68528\\
21945-970 Rio de Janeiro RJ, Brasil.}
\date{\today }
\maketitle

\begin{abstract}
A method of obtaining vector constants of motion for time-independent as
well as time-dependent central fields is discussed. Some well-established
results are rederived in this alternative way and new ones obtained.\ 
\end{abstract}

 \bigskip
\vfill
\noindent $^{\star }$ {e-mail: fsc1omr@fisica.ufsc.br}

\noindent $^{\dagger }$ {e-mail: filadelf@if.ufrj.br}

\noindent $^{\ddagger }$ {e-mail:tort@if.ufrj.br}

\newpage\ 

\section{ Introduction}

It is well known that in classical mechanics the knowledge of all first
integrals of motion of a given problem is equivalent to finding its complete
solution. Nowadays the search for first integrals has assumed an increasing
importance in the determination of the integrability of a dynamical system.
It is extremely important to know if a non-linear dynamical system will
present chaotic behavior in some regions of the phase space. The notion of
integrability is related to the existence of first integrals of motion.
Several methods of finding first integrals are available in the literature
for example, Lie's method \cite{Olver86}, Noether's theorem \cite
{Cantrijn&Sarlet81}, or the direct method \cite{Whittaker37}. Even if not
all first integrals of motion associated with the problem at hand are found,
it may happen that the ones which are obtained contribute to the discovery
of the solution we are seeking for. Nevertheless, if we do find the solution
we are after by solving the equations of motion in a straightforward way, it
still may be profitable to look for additional constants of motion. Such is
the case of the Kepler problem where the knowledge of the Laplace-Runge-Lenz
vector \cite{Goldstein}, \cite{Laplace-Runge-Lenz} allows us to obtain the
orbit in a simple way.

Of the inexhaustible wealth of problems which we can find in classical
mechanics one of the most aesthetically appealing and important is the
central field problem. Energy and angular momentum associated with this type
of field are well known conserved quantities. However, other vector and
tensor conserved quantities have been associated with some particular
central fields. The Laplace-Runge-Lenz vector is a vector first integral of
motion for the Kepler problem; the Fradkin tensor \cite{Fradkin} is
conserved for the case of the harmonic oscillator and for any central field
it is possible to find a vector first integral of motion as was shown in 
\cite{Yan}. In the general case these additional integrals of motion turn
out to be complicated functions of the position $\mathbf{r}$ and linear
momentum $\mathbf{p}$ of the particle probing the central field. When orbits
are closed and degenerated with respect to the mechanical energy, however,
we should expect these additional constant of motion to be simple function
of $\mathbf{r}$ and $\mathbf{p}$. In this article we wish to exploit further
this line of reasoning by determining the existence of such additional
vector first integrals of motion for the time-dependent Kepler and isotropic
harmonic oscillator problems. In particular, we will show that for the
time-dependent Kepler problem the existence of a vector constant of motion
coupled to a simple transformation of variables turns the problem easily
integrable.

The structure of this paper goes as follows: in section 2 we establish the
conditions which guarantee the existence of a vector first integral for a
general central force field. In section 3 we put the method to test by
rederiving some well known results such as the conservation of angular
momentum in an arbitrary central field, the conservation of the
Laplace-Runge-Lenz vector for the Kepler problem, and the conservation of
the Fradkin tensor fixing en route these specific fields to which they
correspond. In section 4 we consider the time-dependent case establishing
generalizations of the examples considered before and presenting new ones.
In section 5 we show that the existence of a vector first integral enable us
to find the orbits of a test particle. This is accomplished for the case of
harmonic oscillator, and the time-dependent Kepler problem. Also the period
of the time-dependent Kepler problem is obtained. Finally, section 6 is
reserved for final comments.

\section{Constructing vector constants of motion}

The force $\mathbf{f}\left( \mathbf{r},t\right) $ acting on a test particle
moving in a central but otherwise arbitrary and possible time-dependent
field of force $g(r,t)$ can be written as

\begin{equation}
\mathbf{f}\left( \mathbf{r},t\right) =g(r,t)\,\,\mathbf{r},
\label{centralforce}
\end{equation}
where $\mathbf{r=r}\left( t\right) $ is the position vector with respect to
the center of force, $r$ is its magnitude, and $t$ is the time. To this test
particle we assume that it is possible to associate a vector\textbf{\ }$%
\mathbf{j}$ which in principle can be written in the form

\begin{equation}
\mathbf{j}\left( \mathbf{p},\mathbf{r},t\right) =A\left( \mathbf{p},\mathbf{r%
},t\right) \,\mathbf{p}+B\left( \mathbf{p},\mathbf{r},t\right) \,\mathbf{r},
\label{jay}
\end{equation}
where $\mathbf{p}=\mathbf{p}\left( t\right) :=m\,\mathbf{\dot{r}}$ $\left(
t\right) $ is the linear momentum, $m$ is the reduced mass and $A$, $B$ are
arbitrary scalar functions of $\mathbf{p},\mathbf{r}$ and $t$. Taking the
total time derivative of (\ref{jay}) and making use of Newton's second law
of motion we readily obtain

\begin{equation}
\frac{d\,\mathbf{j}}{dt}=\left( Ag+\frac{dB}{dt}\right) \mathbf{r}+\left( 
\frac{dA}{dt}+\frac Bm\right) \mathbf{p}.
\end{equation}
If we assume that $\mathbf{j}$ is a constant of motion it follows that the
functions $A$ and $B$ must satisfy

\begin{equation}
Ag+\frac{dB}{dt}=0,  \label{A}
\end{equation}

\begin{equation}
\frac{dA}{dt}+\frac Bm=0.  \label{B}
\end{equation}
Eliminating $B$ between (\ref{A}) and (\ref{B}) we obtain

\begin{equation}
m\frac{d^2A}{dt^2}-gA=0.  \label{A2}
\end{equation}
It follows from (\ref{B}) that\textbf{\ }$\mathbf{j}$ can be written in the
form

\begin{equation}
\mathbf{j}=A\,\mathbf{p}-m\frac{dA}{dt}\,\mathbf{r}.  \label{jay2}
\end{equation}
Therefore, since (\ref{A2}) is equivalent to both (\ref{A}) and (\ref{B}) if
the field $g(r,t)$ is known any solution of (\ref{A2}) will yield a vector
constant of motion of the form given by (\ref{jay2}). Equation (\ref{A2}),
however, is a differential equation whose solution may turn out to be a hard
task to accomplish. Nevertheless, we can make progress if instead of trying
to tackle it directly we make plausible guesses concerning $A$ thereby
linking \textbf{\ }$\mathbf{j}$ to specific forms of the field $g(r,t)$.
This procedure is tantamount to answering the following question: given $%
\mathbf{j}$ what type of central field will admit it as a constant of
motion? The answer is given in the next section.

\section{Simple examples}

With $\mathbf{r}$, $\mathbf{p}$ and a unit constant vector $\mathbf{\hat{u}}$
we can construct the following scalars: $\mathbf{\hat{u}}\cdot \mathbf{r},$ $%
\mathbf{\hat{u}}\cdot \mathbf{p}$ and $\mathbf{r}\cdot \mathbf{p}.$ Other
possibilities will be considered later on. For the moment let us consider
some simple possibilities for the scalar function $A(\mathbf{p},\mathbf{r}%
,t) $.

Consider first $A(\mathbf{p},\mathbf{r},t)=\mathbf{\hat{u}}\cdot \mathbf{r}.$
It is immediately seen that this choice for $A$ satisfies (\ref{A2}) for any
function $g(r,t).$ The constant vector $\mathbf{j}$ reads

\begin{equation}
\mathbf{j=}(\mathbf{\hat{u}\cdot r})\mathbf{\,p-}(\mathbf{\hat{u}\cdot p})%
\mathbf{\,r},  \label{jay3}
\end{equation}
and it can be related to the angular momentum $\mathbf{L}=\mathbf{r}\times \,%
\mathbf{p}$ as follows. Firstly we recast (\ref{jay3}) into the form

\begin{equation}
\mathbf{j}=\mathbf{M\cdot \hat{u}},
\end{equation}
where $\mathbf{M}=\mathbf{p\,r}-\mathbf{r\,p}.$ Since $\mathbf{\hat{u}}$ is
a constant vector we conclude that the constancy of $\mathbf{j}$ is
equivalent to the constancy of $\mathbf{M}$ whose components are $%
M_{jk}=p_jx_k-x_jp_k$, with $i,j,k=1,2,3$. The antisymmetrical tensor $%
\mathbf{M}$ is closely related to angular momentum $\mathbf{L}$ of the test
particle. In fact, it can be easily shown that $-2L_i=\varepsilon
_{ijk}M_{jk}$, where $L_i$ is the i-th angular momentum component and $%
\varepsilon _{ijk}$ is the usual permutation symbol or Levi-Civita density.
Therefore, this simple choice for $A$ leads to conservation of angular
momentum for motion under a central arbitrary field $g(r,t)$.

Consider now the choice $A(\mathbf{p},\mathbf{r},t)=\mathbf{\hat{u}}\cdot 
\mathbf{p}$. Then making use of Newton's second law it follows that (\ref
{jay2}) is satisfied if we find a solution to

\begin{equation}
\frac{d\,g(r,t)}{dt}\,\mathbf{\,\hat{u}}\cdot \mathbf{r}=0.  \label{gee}
\end{equation}
For arbitrary values of $\mathbf{\hat{u}}\cdot \mathbf{r}$ we can find a
solution to (\ref{gee}) if and only if $\stackrel{.}{g(r,t)}\equiv 0,$ or $%
g=g_0$ where $g_0$ is a constant. In this case we can write 
\begin{equation}
\mathbf{j}=(\mathbf{\hat{u}\cdot p})\mathbf{\,p}-g_0\,(\mathbf{\hat{u}}\cdot 
\mathbf{r})\mathbf{\,r}.  \label{jay4}
\end{equation}

If we choose the constant to be equal to $-k$ then the central force field
will correspond to an isotropic harmonic oscillator, $\mathbf{f}=-k\,\mathbf{%
r}.$ As before, (\ref{jay4}) can be recasted into the form

\begin{equation}
\mathbf{j}=2\,m\,\mathbf{F}\cdot \mathbf{\hat{u}},
\end{equation}
where $\mathbf{F}$ is given by

\begin{equation}
\mathbf{F}=\frac{\mathbf{p\,p}}{2m}+k\frac{\mathbf{r}\,\mathbf{r}}2.
\end{equation}
The tensor $\mathbf{F}$ is symmetrical and is known as the Fradkin tensor 
\cite{Fradkin}. Finally, consider $A=\mathbf{r}\cdot \mathbf{p}.$ For this
choice of $A$ (\ref{jay2}) yields

\begin{equation}
\frac 1g\frac{dg}{dt}+\frac 3r\frac{dr}{dt}=0,  \label{gee2}
\end{equation}
where we have made use of (\ref{centralforce}) and also of the fact that $d%
\mathbf{\hat{r}/}dt$ and $\mathbf{r}$ are perpendicular vectors. Equation (%
\ref{gee2}) can be easily integrated if $g$ is considered to be a function
of the radial distance $r$ only. If this is the case we obtain the Kepler
field $g(r)=-k/r^3.$ The constant vector $\mathbf{j}$ is then given by

\begin{equation}
\mathbf{j}=(\mathbf{r}\cdot \mathbf{p})\,\mathbf{p}-(\mathbf{p}\cdot \mathbf{%
p})\,\mathbf{r}+mk\frac{\mathbf{r}}r,  \label{jay5}
\end{equation}
Making use of a well known vector identity we can recast (\ref{jay5}) into
the form,

\begin{equation}
\mathbf{j}=\mathbf{L}\times \mathbf{p}-mk\frac{\mathbf{r}}r.  \label{jay6}
\end{equation}
Therefore $\mathbf{j}$ can be equaled to minus the Laplace-Runge-Lenz vector 
$\mathbf{A}$. From (\ref{jay6}) and the condition $\mathbf{j}\cdot \mathbf{r}%
=-\mathbf{A}\cdot \mathbf{r}=0$ the allowed orbits for the Kepler problem
can be obtained in a simple way, see for example \cite{Goldstein}.

\section{Time-dependent fields}

Let us now consider time-dependent central force fields for which we can
build more general vector first integrals of motion. As with the
time-independent case there are of course several possibilities when it
comes to the choice of a function $A$ for a time-dependent central field.
Here is one

\begin{equation}
A=\phi (t)\,\mathbf{r}\cdot \mathbf{p}+\psi (t)\mathbf{\,r}\cdot \mathbf{r}.
\label{A4}
\end{equation}
Evaluating the second derivative of (\ref{A4}) we obtain

\[
\frac{d^2A}{dt^2}=\left( \frac{d^2\phi }{dt^2}+\frac{4g\phi }m+\frac 4m\frac{%
d\psi }{dt}\right) \mathbf{r}\cdot \mathbf{p}+\left( 2\frac{d\phi }{dt}%
g+\phi \frac{dg}{dt}+\frac{d^2\psi }{dt^2}+\frac{2\psi g}m\right) \mathbf{r}%
\cdot \mathbf{r} 
\]

\begin{equation}
+\frac 2{m^2}\left( m\frac{d\phi }{dt}+\psi \right) \mathbf{p}\cdot \mathbf{p%
}.  \label{DDA}
\end{equation}
Where we have made use of (\ref{centralforce}). If we impose the additional
condition 
\begin{equation}
\psi +m\frac{d\phi }{dt}=0,  \label{PSIPHI}
\end{equation}
we eliminate the quadratic term in \textbf{\ }$\mathbf{p}$. With the
condition given by (\ref{PSIPHI}) we can substitute for $\psi $ in (\ref{A4}%
) and (\ref{DDA}) and take the results into (\ref{A2}) thus obtaining

\begin{equation}
3\left( -m\frac{d^2\phi }{dt^2}+g\phi \right) \mathbf{r}\cdot \frac{d\,%
\mathbf{r}}{dt}+\left( \phi \frac{dg}{dt}-m\frac{d^3\phi }{dt^3}+g\frac{%
d\phi }{dt}\right) \mathbf{r}\cdot \mathbf{r}=0.  \label{PSIPHI2}
\end{equation}
Equation (\ref{PSIPHI2}) can be rewritten as

\begin{equation}
\frac 32\left( -m\frac{d^2\phi }{dt^2}+g\phi \right) \frac{d(\mathbf{r}^2)}{%
dt}+\left[ \frac d{dt}\left( -m\frac{d^2\phi }{dt^2}+g\phi \right) \right] 
\mathbf{r}^2=0,
\end{equation}
and easily integrated so as to yield

\begin{equation}
g(r,t)=\frac m\phi \frac{d^2\phi }{dt^2}+\frac C{\phi r^3}.  \label{gee3}
\end{equation}
The vector first integral of motion associated with (\ref{gee3}) is

\begin{equation}
\mathbf{j}=\left( \phi \,\mathbf{r}\cdot \mathbf{p}-m\frac{d\phi }{dt}\,%
\mathbf{r}\cdot \mathbf{r}\right) \,\mathbf{p}+\left( m\frac{d\phi }{dt}\,%
\mathbf{r\cdot p}-\phi \,\mathbf{p\cdot p}-\frac{mC}r\right) \mathbf{r},
\end{equation}
which can be simplified and written in the form

\begin{equation}
\mathbf{j}=m\phi ^2\mathbf{L}\times \frac d{d\,t}\left( \frac{\mathbf{r}}%
\phi \right) -\frac{mC\,\mathbf{r}}r.  \label{jay7}
\end{equation}
where we have made used of the fact the angular momentum is constant for any
arbitrary central field whether it is time-independent or not. If in (\ref
{jay7}) we set $\phi =1$ and $C\neq 0$, then from (\ref{gee3}) we see that $%
g\left( r,t\right) $ is the Kepler field and $\mathbf{j}$ is minus the
Laplace-Runge-Lenz vector as before; the scalar function $A\left( \mathbf{r},%
\mathbf{p},t\right) $ reduces to $\mathbf{r}\cdot \mathbf{p}$ which we have
already employed in section 3. If we set $(m/\phi )\,d^2\phi /dt^2=-k(t),$
that is, if $\phi $ is an arbitrary function of the time, and also $C=0$, we
have the time-dependent isotropic harmonic oscillator field, $\mathbf{F}%
\left( r\right) =-k(t)\,\mathbf{r}$. In this case $\mathbf{j}$ is equal to
the first term on the R.H.S. of (\ref{jay7}). If $(m/\phi )d^2\phi /dt^2=-k$
and $C=0$, we have the time-independent isotopic harmonic oscillator field
but this time $\mathbf{j}$ is not the same vector as the one we have
obtained before. The reason for this is our choice (\ref{A4}) for the scalar
function $A\left( \mathbf{r},\mathbf{p},t\right) $ which is not reducible to
the form $\mathbf{\hat{u}\cdot p}$ employed previously. 

As a last example let us consider again the time-dependent isotropic
harmonic oscillator and show how it is possible to generalize the Fradkin
tensor for this case. Let the function $A\left( \mathbf{r},\mathbf{p}%
,t\right) $ be written as

\begin{equation}
A=\phi (t)\,\mathbf{\hat{u}\cdot r}+\psi (t)\,\mathbf{\hat{u}\cdot p}.
\label{A5}
\end{equation}
The first and the second derivative of $A$ read

\begin{equation}
\frac{dA}{dt}=\frac{d\phi }{dt}\,\mathbf{\hat{u}\cdot r}+\frac \phi m\,%
\mathbf{\hat{u}\cdot p}+\frac{d\psi }{dt}\,\mathbf{\hat{u}\cdot p}+g\,\psi \,%
\mathbf{\hat{u}\cdot r},  \label{DA5}
\end{equation}
and 
\begin{equation}
\frac{d^2A}{dt^2}=\left( \frac{d^2\phi }{dt^2}+2g\frac{d\psi }{dt}+\frac{dg}{%
dt}\;\psi +\frac{g\,\phi }m\right) \mathbf{\hat{u}\cdot r+}\left( \frac{%
d^2\psi }{dt^2}+2g\frac{d\phi }{dt}+\frac{g\,\psi }m\right) \mathbf{\hat{u}%
\cdot p}  \label{DDA5}
\end{equation}
Taking (\ref{DDA5}) into (\ref{A2}) we obtain the condition

\begin{equation}
m\left( \frac{d^2\phi }{dt^2}+2g\,\frac{d\psi }{dt}+\frac{dg}{dt}\,\psi
\right) \mathbf{\hat{u}\cdot r}+\left( 2\,\frac{d\phi }{dt}+m\,\frac{d^2\psi 
}{dt^2}\right) \mathbf{\hat{u}\cdot p}=0.  \label{CON}
\end{equation}
Imposing the additional condition

\begin{equation}
2\frac{d\phi }{dt}+m\frac{d^2\psi }{dt^2}=0,  \label{CON2}
\end{equation}
(\ref{CON}) becomes

\begin{equation}
m\frac{d^3\psi }{dt^3}-4g\frac{d\psi }{dt}-2\frac{dg}{dt}\psi =0.
\label{CON3}
\end{equation}
We can solve (\ref{CON3}) thoroughly if $g\left( r,t\right) $ is a function
of the time $t$ only. In this case, as before, we end up with the
time-dependent isotropic harmonic oscillator. The vector $\mathbf{j}$
associated with (\ref{A5}) can be obtained as follows: first we integrate (%
\ref{CON2}) thus obtaining

\begin{equation}
\phi =-\frac m2\frac{d\psi }{dt}+C,  \label{PHY}
\end{equation}
where $C$ is an integration constant. Then making use of (\ref{A5}), (\ref
{PHY}) and (\ref{DA5}) we arrive at

\begin{eqnarray}
\mathbf{j} &=&\left[ \left( -\frac m2\frac{d\psi }{dt}+C\right) \mathbf{\hat{%
u}\cdot r}+\psi \mathbf{\,\hat{u}\cdot p}\right] \,\mathbf{p} \\
&&+\left[ \left( \frac{m^2}2\frac{d^2\psi }{dt^2}-mg\psi \right) \mathbf{%
\hat{u}\cdot r}-\left( \frac m2\frac{d\psi }{dt}+C\right) \mathbf{\hat{u}%
\cdot p}\right] \,\mathbf{r},  \nonumber
\end{eqnarray}
or in terms of components

\begin{equation}
j_i=F_{ij}u_j,
\end{equation}
where $F_{ij}$ is defined by

\begin{eqnarray}
F_{ij} &=&\left( -\frac m2\frac{d\psi }{dt}+C\right) p_ix_j+\psi p_ip_j
\label{FRAD2} \\
&&+\left( \frac{m^2}2\frac{d^2\psi }{dt^2}-mg\psi \right) x_ix_j-\left(
\frac m2\frac{d\psi }{dt}+C\right) x_ip_j.  \nonumber
\end{eqnarray}
The constant $C$ in (\ref{FRAD2}) can be made zero without loss of
generality. A generalized Fradkin tensor can now be defined by

\begin{equation}
F_{ij}=\left( -\frac m2\frac{d\psi }{dt}\right) p_ix_j+\psi \,p_ip_j+\left( 
\frac{m^2}2\frac{d^2\psi }{dt^2}-mg\psi \right) x_ix_j-\frac m2\frac{d\psi }{%
dt}x_ip_j\ .  \label{GFRAD}
\end{equation}
From (\ref{GFRAD}) we can read out the diagonal components of the
generalized Fradkin tensor

\begin{equation}
F_{ii}=-m^2\frac{d\,\psi }{dt}\;x_i\stackrel{.}{\frac{dx_i}{dt}}+\,\psi
\,p_i^2+\left( \frac{m^2}2\frac{d^2\psi }{dt^2}-mg\right) x_i^2\ .
\end{equation}
It is not hard to see that the trace of this generalized Fradkin tensor $%
\mathbf{F}\left( \mathbf{r},\mathbf{p},t\right) $ becomes the energy of the
particle when $g\left( r,t\right) $ is a constant field.

\section{Obtaining explicit solutions: An alternative way}

Now we wish to show how to take advantage of the vector constant $\mathbf{j}$
to obtain the solution for the Kepler and the isotropic harmonic oscillator
potentials. But firstly we must establish some very general relationships
between the sought for solution $r\left( t\right) $ and $A\left( \mathbf{r},%
\mathbf{p},t\right) $ and $\mathbf{j}$. First notice that (\ref{gee}) can be
recasted into the form

\begin{equation}
\mathbf{j}=m\,\,A\left( \mathbf{r},\mathbf{p},t\right) ^2\frac d{dt}\left[ 
\frac{\mathbf{r}}{A\left( \mathbf{r},\mathbf{p},t\right) }\right] \ ,
\label{jay8}
\end{equation}
where it must be kept in mind that $A\left( \mathbf{r},\mathbf{p},t\right) $
satisfies (\ref{A2}). As we have seen before in specific examples the form
of the vector constant $\mathbf{j}$ depends on the force acting on the
particle. Integrating (\ref{jay8}) we readily obtain

\begin{equation}
\frac{\mathbf{r}\left( t\right) }{A\left( \mathbf{r},\mathbf{p},t\right) }-%
\frac{\mathbf{r}\left( 0\right) }{A\left( \mathbf{r},\mathbf{p},0\right) }=%
\frac{\mathbf{j}}m\int_0^t\frac{d\tau }{A\left( \mathbf{r},\mathbf{p},\tau
\right) ^2}\ .  \label{deltar}
\end{equation}
Equation (\ref{deltar}) can be given a simple but interesting geometrical
interpretation. Assume that the initial conditions $\mathbf{r}\left(
0\right) $ and $\mathbf{p}\left( 0\right) $ are known and therefore the
function $A\left( \mathbf{r},\mathbf{p},0\right) $ can be determined. The
vector $\mathbf{r}\left( 0\right) /A\left( \mathbf{r},\mathbf{p},0\right) $%
is therefore a constant and completely determined vector. As time increases,
the R.H.S. of (\ref{deltar}) increases. The vector on the right side of (\ref
{deltar}) though varying in time has a fixed direction which is determined
by $\mathbf{j}$. Therefore, $\mathbf{r}\left( t\right) /A\left( \mathbf{r},%
\mathbf{p},t\right) $ must increase in order to close the triangle whose
sides are the three vectors involved in (\ref{deltar}). If the orbit is
unlimited then it is easy to see that the following property ensues: there
is an asymptote if in the limit $t\to \infty $\ the definite integral $%
\int_0^t\frac{dt}{A^2}$ is constant. On the other hand, if the orbit is
limited, but not necessarily closed, there will be a position vector $%
\mathbf{r}$ whose direction is parallel to that of the vector $\mathbf{j}$
at the instant $t^{*}$. If the length of the position vector $\mathbf{r}$ is
finite, we can conclude that at the same instant $t^{*}$ the function $%
A\left( \mathbf{r},\mathbf{p},0\right) $ must be zero. Thus, we can see that
the vector $\mathbf{r}\left( t\right) /A\left( \mathbf{r},\mathbf{p}%
,t\right) $ must be reversed at this instant and its evolution is determined
by the fact that its end is on the straight line that contains $\mathbf{j}$.
For $t=t^{*}+\epsilon $, where $\epsilon $ is a positive infinitesimal
number, the vector $\mathbf{r}\left( t^{*}+\epsilon \right) /A\left( \mathbf{%
r},\mathbf{p},t^{*}+\epsilon \right) $ changes its direction abruptly, so to
speak, as shown in the figure, hence in the transition $A\left( \mathbf{r},%
\mathbf{p},t^{*}\right) \rightarrow A\left( \mathbf{r},\mathbf{p}%
,t^{*}+\epsilon \right) $ the scalar function must change its sign.

Let us obtain the solution $\mathbf{r}\left( t\right) $ for the case of the
isotropic harmonic oscillator. A particular solution of (\ref{A4}) for $%
g=-\kappa ,$ where $\kappa $ is the elastic cons$\tan $t is given by

\begin{equation}
A(t)=\cos (\omega t)\ ,  \label{HOSSol1}
\end{equation}
where $\omega =\sqrt{\frac km}$ is the angular frequency. The solution given
by (\ref{HOSSol1}) allows us to write

\begin{equation}
\frac{\ \mathbf{r}\left( t\right) }{\cos (\omega t)}-\mathbf{r}\left(
0\right) =\frac{\ \mathbf{p}\left( 0\right) }m\int_0^t\frac{d\tau }{\cos
^2(\omega \tau )}\ 
\end{equation}
The integral can be readily performed and after some simplifications we
finally obtain

\begin{equation}
\mathbf{r}\left( t\right) =\cos \omega t\ \mathbf{r}\left( 0\right) +\frac
1{m\omega }\sin \omega t\ \mathbf{p}\left( 0\right) \ .
\end{equation}
Therefore we have obtained the solution of the time-independent harmonic
oscillator in an alternative way from the knowledge of the initial
conditions as it should be. Notice that the general solution  $A(t)=A\left(
0\right) \cos \left( \omega t+\theta \right) $ would lead to the same
general result.  Another possible solution in the case of the
time-independent isotropic harmonic oscillator is given by

\begin{equation}
A\left( \mathbf{r},\mathbf{p},t\right) =\mathbf{\hat{u}\cdot p}\left(
t\right) ,
\end{equation}
as can be shown by substituting this solution into (\ref{A2}). This solution
shows that the trajectories have no asymptotes.

Let us now show how we can obtain the orbits in the case of the
time-dependent Kepler problem. Let us begin by rewriting (\ref{jay8}) in
polar coordinates on the plane. In these coordinates the angular momentum
conservation law is written in the form

\begin{equation}
l=mr^2\frac{d\theta }{dt}  \label{ANGMOM}
\end{equation}
and this allows to rewrite (\ref{jay8}) as

\begin{equation}
\mathbf{j}=mA^2\frac d{d\theta }\left( \frac{\mathbf{r}}A\right) \frac{%
d\theta }{dt}=l\left( \frac Ar\right) ^2\frac d{d\theta }\left( \frac{%
\mathbf{r}}A\right) \ .
\end{equation}
Introducing the unitary vectors $\stackrel{\wedge }{\mathbf{r}}$ and $%
\stackrel{\wedge }{\mathbf{\theta }}$ we can write the above equation as

\begin{equation}
\mathbf{j}=l\left[ -\frac d{d\theta }\left( \frac Ar\right) \stackrel{\wedge 
}{\mathbf{r}}+\frac Ar\stackrel{\wedge }{\mathbf{\theta }}\right] \ .
\end{equation}
The components of the vector $\mathbf{j}$ in the direction of $\stackrel{%
\wedge }{\mathbf{r}}$ and $\stackrel{\wedge }{\mathbf{\theta }}$ are given by

\begin{equation}
\mathbf{j}\cdot \stackrel{\wedge }{\mathbf{\theta }}=l\frac Ar\ ,
\label{jtheta}
\end{equation}
and

\begin{equation}
\mathbf{j}\cdot \stackrel{\wedge }{\mathbf{r}}=-l\frac d{d\theta }\left(
\frac Ar\right) \ .  \label{jerr}
\end{equation}
Equation (\ref{jerr}) can be obtained from (\ref{jtheta}) and therefore it
is redundant. In section 4 we determined a generalized Laplace-Runge-Lenz
vector for the time-dependent Kepler problem. The scalar function $A\left( 
\mathbf{r},\mathbf{p},t\right) $ associated with this vector was found to be

\begin{equation}
A=\phi (t)\;\,\mathbf{r\cdot p}-m\frac{d\phi }{dt}r^2\ .  \label{AK}
\end{equation}
Making use of (\ref{ANGMOM}) we can rewrite the linear momentum as a
function of $\theta $ as follows

\begin{equation}
\mathbf{p}=m\frac{d\mathbf{r}}{dt}\frac{d\theta }{dt}=\frac l{r^2}\frac{d%
\mathbf{r}}{d\theta }\ .  \label{PTHETA}
\end{equation}
Taking (\ref{PTHETA}) into (\ref{AK}) and considering $A$ as a function of $%
\theta $ we obtain

\begin{equation}
\frac Ar=-l\frac d{d\theta }\left( \frac \phi r\right) \ .  \label{AR}
\end{equation}
Equations. (\ref{jtheta}) and (\ref{AR}) lead to

\begin{equation}
\frac d{d\theta }\left( \frac \phi r\right) =-\frac{\stackrel{}{\mathbf{j}%
\cdot \stackrel{\mathbf{\wedge }}{\mathbf{\theta }}}}{l^2}=\frac j{l^2}\sin
(\theta -\alpha )\ ,  \label{alpha}
\end{equation}
where $\alpha $ is the angle between $\mathbf{j}$ and the $\mathcal{OX}$
axis (see figure 2). In order to integrate (\ref{alpha}) we assume that the
initial conditions at $t=0$ are known vector functions, i.e.

\begin{equation}
\mathbf{r}(0)=\mathbf{r}_0\ ;
\end{equation}
and

\begin{equation}
\mathbf{p}(0)=\mathbf{p}_0\ .
\end{equation}
In terms of polar coordinates these initial conditions are written as

\begin{equation}
r(\theta _0)=r_0\ ,
\end{equation}
and making use of (\ref{PTHETA})

\begin{equation}
\left( \frac{d\mathbf{r}}{d\theta }\right) _{\theta =\theta _0}=\frac{r_0^2}l%
\mathbf{p}_0\ .
\end{equation}
Upon integrating (\ref{alpha}) we find

\begin{equation}
\frac \phi r=\frac{\phi _0}{r_0}+\frac j{l^2}\left[ \cos (\theta _0-\alpha
)-\cos (\theta -\alpha )\right] \ .  \label{alpha2}
\end{equation}
For the usual time-independent Kepler problem, $\phi =1$ and in this case (%
\ref{alpha2}) takes the form

\begin{equation}
\frac 1r=\frac 1{r_0}+\frac j{l^2}\left[ \cos (\theta _0-\alpha )-\cos
(\theta -\alpha )\right] \   \label{alpha3}
\end{equation}
The scalar product between $\mathbf{j}$ as given by (\ref{jay6}) and $%
\mathbf{r}_0$ permit us to eliminate $\cos (\theta _0-\alpha )$ and leads to
the usual orbit equation

\begin{equation}
\frac 1r=-\frac{mC}{l^2}\left[ 1+\frac j{mC}\cos (\theta -\alpha )\right] \ .
\label{alpha3}
\end{equation}
If we define a new position vector $\mathbf{r}^{\prime }$ according to

\begin{equation}
\mathbf{r}^{\prime }:=\frac{\mathbf{r}}\phi \ ,  \label{RPRIME}
\end{equation}
and redefine our time parameter according to

\begin{equation}
dt^{\prime }:=\frac{dt}{\phi ^2}\ ,  \label{TPRIME}
\end{equation}
we can recast the equation of motion for the time-dependent Kepler problem,
namely

\begin{equation}
m\frac{d^2\mathbf{r}}{dt^2}=\left( \frac m\phi \frac{d^2\phi }{dt^2}+\frac
C{\phi \,r^3}\right) \mathbf{r}  \label{eqmotion}
\end{equation}
into a simpler form. According to (\ref{RPRIME}) and (\ref{TPRIME}) the
velocity and the acceleration transform in the following way

\begin{equation}
\frac{d\mathbf{r}}{dt}=\mathbf{r}^{\prime }\frac{d\phi }{dt}+\frac 1\phi 
\frac{d\mathbf{r}}{dt^{\prime }}^{\prime }  \label{vel}
\end{equation}
and

\begin{equation}
\frac{d^2\mathbf{r}}{dt^2}=\mathbf{r}^{\prime }\frac{d^2\phi }{dt^2}+\frac
1{\phi ^3}\frac{d^2\mathbf{r}}{dt^{\prime \,2}}^{\prime }\ .  \label{accel}
\end{equation}
where we have taken advantage of the fact that $\mathbf{r},\mathbf{r}%
^{\prime }$ and $\phi $ can be considered as functions of $t$ or $t^{\prime
} $. Making use of (\ref{accel}) the equation of motion (\ref{eqmotion}) can
be written as

\begin{equation}
m\frac{d^2\mathbf{r}}{dt^{\prime 2}}^{\prime }=\frac C{(r^{\prime })^3}%
\mathbf{r}^{\prime }\ .  \label{eqmotion2}
\end{equation}
Equation (\ref{eqmotion2}) corresponds to the usual time-independent Kepler
problem whose solution is given by (\ref{alpha3}). Equations (\ref{RPRIME})
and (\ref{TPRIME}) show that the open solutions of (\ref{eqmotion2}) are
transformed into the open solutions of (\ref{eqmotion}) with the same
angular size and that closed solutions of (\ref{eqmotion2}) are associated
with spiraling solutions of (\ref{eqmotion}). The period of the orbit of (%
\ref{eqmotion2}) is related to the time interval that the spiraling particle
takes to cross a fixed straight line. Representing this time interval by $%
T_0 $ we have

\begin{equation}
T=\int_0^{T_0}\frac{dt}{\phi ^2}\ .  \label{T}
\end{equation}
As an application of the above remarks suppose we are looking for the form
of the function $\phi $ which yields a circular orbit with radius $R$ as a
solution to (\ref{eqmotion})? To find this function we see from (\ref
{eqmotion}) that we have to solve the following equation differential
equation

\begin{equation}
\frac{d^2\phi }{dt^2}+\frac{\left| C\right| }{mR^3}\phi =\frac{\left|
C\right| }{mR^3}.
\end{equation}
The solution is

\begin{equation}
\phi \left( t\right) =1+\phi _0\cos \left( \omega \,t+\beta \right) \ ,
\end{equation}
where $\phi _0$ is a constant and $\omega =\sqrt{\frac{\left| C\,\right| }{%
mR^3}}$ and $\beta $ an arbitrary phase angle . The constant $\phi _0$ may
be chosen so that the transformed solution will be a given ellipse as we
show below. Equation (\ref{RPRIME}) leads to

\[
\frac 1{r^{\prime }}=R^{-1}\left[ 1+\phi _0\cos \left( \omega \,t+\beta
\right) \ \right] \ . 
\]
Comparing with (\ref{alpha3}) we obtain

\begin{equation}
R=\frac{l^2}{m\,\left| C\right| }\ ,
\end{equation}
and

\begin{equation}
\phi _0=-\frac j{m\,\left| C\right| }.
\end{equation}
The period of this circular orbit is given

\begin{equation}
T_0=\frac{2\pi }\omega =2\pi \sqrt{\frac{mR^3}{\left| C\,\right| }}.
\end{equation}
and using (\ref{T}) we get

\begin{equation}
T=\int_0^{\frac{2\pi }\omega }\frac{dt}{(1+e\cos \omega t)^2}=\frac{2\pi }%
\omega \frac 1{(1-e^2)^{\frac 32}}\ 
\end{equation}
where $e=\left| \phi _0\right| $ is the eccentricity. Making use of $%
1-e^2=R/a$, where $a$ is the major semi-axis we finally obtain the orbital
period

\begin{equation}
T=2\pi a^{\frac 32}\sqrt{-\frac mC\ .}
\end{equation}
To conclude consider the total mechanical energy associated with (\ref
{eqmotion2}) 
\begin{equation}
E=\frac{\mathbf{p}^{\prime \,2}}{2\,m}+\frac C{\ r^{\prime }}=const.
\end{equation}
Since $\mathbf{p}^{\prime }$ and $\mathbf{p}$ are related by 
\begin{equation}
\mathbf{p}^{\prime }=\phi \left( t\right) \,\mathbf{p}-m\,\dot{\phi}\left(
t\right) \,\mathbf{r}
\end{equation}
and $r^{\prime }$ and $r$ by (\ref{RPRIME}) we easily obtain

\begin{equation}
E=\phi ^2\frac{\mathbf{p}^2}{2m}-2\phi \,\frac{d\phi }{dt}\mathbf{r\cdot p}%
+\left( \frac{d\phi }{dt}\right) ^2\frac{r^2}2+C\frac \phi r
\end{equation}
which is a conserved quantity and can be interpreted as a generalization of
the energy of the particle under the action of a time-dependent Kepler field.

\section{Laplace-Runge-Lenz type of vector constants for arbitrary central
fields}

Equation (\ref{gee3}) determines a time-dependent Kepler field $g\left(
r,t\right) ,$ where the variables $r$ and $t$ are independent. If, however,
we consider the orbit equation $r\left( t\right) $ we can eliminate the time
variable and define the function $g\left( r,t\left( r\right) \right) $ which
can be understood as an arbitrary function of $r.$ For the sake of
simplicity we denote this function by $g\left( r\right) $. The function $%
\phi \left( t\right) $ which transforms the Kepler problem  when understood
as a function of $r$ transforms the Kepler field in an arbitrary central
field. Let us write Eq. (\ref{gee3}) in the form
\begin{equation}
m\frac{d^2\phi }{dt^2}-g\left( r,t\right) \phi +\frac C{r^3}=0
\label{gee3prime}
\end{equation}
and consider the transformation 
\begin{equation}
\frac{d^2\phi }{dt^2}=\frac{d^2\phi }{dr^2}\;\left( \frac{dr}{dt}\right) ^2+%
\frac{d\phi }{dr}\,\frac{d^2r}{dt^2}
\end{equation}
Energy conservation and the equation of motion allow us to write 
\begin{equation}
\left( \frac{dr}{dt}\right) ^2=\frac 2m\left[ E-V\left( r\right) -\frac{l^2}{%
2mr^2}\right] 
\end{equation}
and 
\begin{equation}
\frac{d^2r}{dt^2}=\frac{rg\left( r\right) }m+\frac{l^2}{m^2r^3}
\end{equation}
where $E$ is the energy of the particle and $l$ its angular momentum. Taking
these three last equations into account Eq. (\ref{gee3prime}) reads now 
\begin{equation}
2\left[ E-V\left( r\right) -\frac{l^2}{2mr^2}\right] \frac{d^2\phi }{dr^2}%
+\left[ rg\left( r\right) +\frac{l^2}{m^2r^3}\right] \frac{d\phi }{dr}%
-g\left( r\right) \phi +\frac C{r^3}=0  \label{FR}
\end{equation}
Equation (\ref{FR}) permit us to determine the function $\phi \left(
r\right) $ for any potential $V\left( r\right) $. Thus, we conclude that a
central field problem can be transformed into a time-dependent Kepler
problem. When $g\left( r\right) $ describes the Kepler field the solution of
(\ref{FR}) is $\phi \left( r\right) =1$. Sometimes it is convenient to
perform a second change of variables by defining the transformation 
\begin{equation}
\phi =\psi -\frac{mC}{l^2}
\end{equation}
Then Eq. (\ref{FR}) becomes 
\begin{equation}
2\left[ E-V\left( r\right) -\frac{l^2}{2mr^2}\right] \frac{d^2\psi }{dr^2}%
+\left[ \frac{rg\left( r\right) }m+\frac{l^2}{m^2r^3}\right] \frac{d\psi }{dr%
}-g\left( r\right) \psi =0  \label{FRprime}
\end{equation}
As an example consider $V\left( r\right) =k/r$. Then the solution of (\ref
{FRprime}) is simply 
\begin{equation}
\psi \left( r\right) =c_1\left( kmr+l^2\right) +c_2\sqrt{l^2+2kmr-2mEr^2}
\end{equation}

\section{Conclusions}

In this paper we have outlined a simple and effective method for treating
problems related with time-dependent and time independent central force
fields. In particular we have dealt with the Kepler problem and the
isotropic harmonic oscillator fields. We have been able to rederive some
known results from an original point of view and generalize others. The
central force field has been discussed in the literature from many points of
view. The difficulty in finding vector constants of motion for central
fields stem from the fact that in general orbits for these type of problem
are not closed, therefore any new ways to attack those problems are
welcome.. In our method this difficulty is transferred, so to speak, to the
obtention for each possible central field, which can be time-dependent or
not, of a certain scalar function of the position, linear momentum and time.
For a given central field this scalar function is a solution of (\ref{A2}).
In the general case, the obtention of the scalar function is a difficult
task. Judicious guesses, however, facilitate the search for solution of (\ref
{A2}) and this is what we have done here.

\end{document}